\documentclass{article}

     \usepackage[preprint]{neurips_2019}

\usepackage[utf8]{inputenc} 
\usepackage[T1]{fontenc}    
\usepackage{hyperref}       
\usepackage{url}            
\usepackage{booktabs}       
\usepackage{amsfonts}       
\usepackage{nicefrac}       
\usepackage{microtype}      

\def\PP{{{\rm l}\kern - .15em {\rm P} }}
\def\PN2{{\PP_{N}-\PP_{N-2}}}

\newcommand{\bA}{\boldsymbol{A}}

\newcommand{\be}{\boldsymbol{e}}

\newcommand{\bFF}{{\boldsymbol F}}

\newcommand{\bW}{\boldsymbol{W}}

\newcommand{\bX}{\boldsymbol{X}}

\newcommand{\deleted}[1]{{}}

\usepackage{bigints}
\usepackage{graphicx}
\graphicspath{{./figs/}}
\title{Analytic Continuation of Noisy Data Using Adams Bashforth ResNet}

\author{%
  Xuping Xie \\
Computation and Applied Mathematics\\
  Oak Ridge National Laboratory\\
  \texttt{xiex@ornl.gov} \\
  \And
  Feng Bao\\
  Department of Mathematics\\
  Florida State University\\
  \texttt{bao@math.fsu.edu}
   \And
   Thomas Maier \\
   Center for Nanophase Materials Sciences\\
   Oak Ridge National Laboratory \\
   \texttt{maierta@ornl.gov} \\
   \And
   Clayton Webster \thanks{The author is also affiliated with Computational and Applied Mathematics, Oak Ridge National Laboratory}\\
Department of Mathematics\\
University of Tennessee, Knoxville\\
   \texttt{cwebst13@utk.edu} 
}

\begin{document}

\maketitle

\begin{abstract}
 We propose a data-driven learning framework for the analytic continuation problem in  numerical quantum many-body physics. Designing an accurate and efficient framework for the analytic continuation of imaginary time using computational data is a grand challenge that has hindered meaningful links with experimental data. The standard Maximum Entropy (MaxEnt)-based method is limited by the quality of the computational data and the availability of prior information. Also, the MaxEnt is not able to solve the inversion problem under high level of noise in the data. Here we introduce a novel learning model for the analytic continuation problem using a Adams-Bashforth residual neural network (AB-ResNet). The advantage of this deep learning network is that it is model independent and, therefore, does not require prior information concerning the quantity of interest given by the spectral function. More importantly, the ResNet-based model achieves higher accuracy than MaxEnt for data with higher level of noise. Finally, numerical examples show that the developed AB-ResNet is able to recover the spectral function with accuracy comparable to MaxEnt where the noise level is relatively small.
 
\end{abstract}

\section{Introduction}
\label{sec:intro}

Quantum Monte Carlo (QMC) methods are widely used to study the finite temperature physics of strongly interacting electron systems. The underlying algorithms are generally formulated on the imaginary time axis to treat the finite temperature dynamics of the many-body system. To extract the real time dynamics, an additional analytic continuation of the imaginary time $\tau$ data to the real time or frequency $\omega$-axis is required to extract the quantity of interest. This process is a highly ill-conditioned inverse problem so that small perturbations of the input data result in large uncertainties in the resulting spectral function $A(\omega)$. The challenge is rooted in the integral equation
\begin{equation}\label{eq:AC}
    G(\tau) = \int_{-\infty}^\infty \frac{e^{-\tau\omega}}{1+e^{-\beta\omega}} A(\omega) d\omega\,.
\end{equation}
Here $G(\tau)$ is the imaginary time QMC data for a fermionic observable such as the single-particle Green's function, $K(\tau,\omega) = \exp(-\tau\omega)/(1+\exp(-\beta\omega))$ is the kernel function with $\beta=1/T$ the inverse temperature, and $A(\omega)$ is the quantity of interest. The process of inverting this equation is numerically unstable because of the exponentially small tails in the kernel function for large $\omega$, and is especially sensitive to the Monte Carlo sampling error in $G(\tau)$ \cite{levy2017implementation}. 

\paragraph{Related Work} Several approaches have been proposed to address the analytic continuation problem. The most commonly used framework based on Bayesian inference is the MaxEnt method \cite{Gull:1984jk}, pioneered in the works \cite{Jarrell:1996uo,Silver:1990eb} for the analytic continuation problem given by Eq.~(\ref{eq:AC}). The MaxEnt method regularizes the inversion problem through the introduction of an entropy-like term that measures the deviation from a default spectrum and then determines the most probable spectrum $A(\omega)$ using deterministic optimization. A related method that uses consistent constraints for the regularization was introduced in \cite{Prokofev:2013hi}. Both methods have the drawback that prior information about the possible spectrum $A(\omega)$ is needed for the regularization.

An alternative idea, which in principle does not rely on prior information is based on stochastic optimization. The work \cite{Sandvik:1998jr} uses Monte Carlo sampling of possible spectra weighted by Boltzmann weights with a fictitious temperature. This method was later related to MaxEnt in a certain limit in the paper \cite{Beach:04,Fuchs:2010it}. The effort \cite{Fuchs:2010it} showed that Bayesian inference can be used to eliminate the ficticious temperature. Moreover, the work \cite{Mishchenko:2000co} developed a  stochastic optimization based method to randomly sample possible optimal solutions $A(\omega)$, which implicitly regularizes the problem by allowing less optimal solutions. A more accessible and less complex variant of this approach that uses a Gaussian process for implicit regularization was recently introduced in \cite{Bao:2016ca} and shown to provide spectra similar to MaxEnt. 


\paragraph{Our approach} Much recent work has been proposed on the mathematical connections between ResNet and differential equations, see, e.g., \cite{beck2017machine,lite19jmlr,long18a,ma2018model}. The work \cite{chen2018neural} introduced ODE-net which parametrize the derivative of the hidden state using deep ResNet. Other efforts \cite{chang2018reversible,chang2017multi,haber2017stable,lu2017beyond} proposed the dynamic system view of ResNet and provide connections between numerical ODE and deep ResNet architecture. Moreover, the functional approximation ability of ResNet has also been explored \cite{wang2018exponential,wang2018deep,wume18nips}.
The work \cite{lin2018resnet} proved ResNet can be considered as a universal approximator with one hidden layer and has certain advantages to fully connected neural networks. The paper \cite{fournier2018artificial} used a simple artificial neural network to approximate the kernel of the inversion and \cite{fuchs2010analytic} introduced stochastic inference approach for this problem \cite{arsenault2016projected,jarrell1996bayesian,yoon2018analytic}.
Motivated by the recent development of residual networks \cite{he2016deep}, we propose a Adams-Bashforth residual network architecture to generate a more stable inversion of the kernel under high noise data for the analytic continuation problem.

This paper is strucuted as follows: In Section~\ref{sec:maxent}, we briefly introduce the classical MaxEnt method. In Section~\ref{sec:ab-resnet}, we present the recent mathematical interpretation of ResNet and our new network architecture. In Section \ref{sec:numerical}, we demonstrate the effectiveness of our method using numerical experiments. In Section \ref{sec:conclu}, we discuss some further works need to investigate for our model.

\section{Maximum Entropy}
\label{sec:maxent}
The MaxEnt method, based on Bayesian inference, is based on the following formula, $P(A|G)=P(G|A)P(A)$, where $P(A|G)$ is the posterior probability of the spectrum $A(\omega)$, given the data $G(\tau)$, the prior probability $P(A)$ contains prior information about spectral function, and the likelihood function $P(G|A)$ measures the quality of the fit between $G(\tau)$ and $KA$, where $K$ is kernel defined above. The problem of finding the most probable spectrum $A$ given the data $G$ is thereby converted into the much easier problem of optimizing the likelihood function and prior probability. The likelihood is defined according to central limit theorem as, $P(G|A) = exp^{-\chi^2[A]/2}$. For the samples $A^i$ and the corresponding $G^i(\tau)$,
\begin{eqnarray}
\chi^2[A^i] = \sum_{n,m=1}^M(G^i(\tau_m)-G(\tau_n))C^{-1}_{mn}(G^i(\tau_m)-G(\tau_n)),
\end{eqnarray}
with $C_{mn}$ being the covariance matrix. The MaxEnt method uses least-square to minize $\chi^2$ with Kullback-Leibler (KL) divergence as the regularization $S[A]$, namely:
\begin{eqnarray}
S[A] = -\int d\omega A(\omega)\ln\biggl(\frac{A(\omega)}{d(\omega)}\biggr).
\end{eqnarray}
Finally, the MaxEnt method minizes the function $Q = \frac{1}{2}\chi^2[A]-\alpha S[A]$. $\alpha$ is a control parameter for the regularization. In our numerical tests, the MaxEnt results are obtained by averaging over $\alpha$ the optimal spectrum $A_{\alpha}$ for each $\alpha$.
\section{Adams-Bashforth (AB) Residual Network}
\label{sec:ab-resnet}

\subsection{ODE representation of ResNet}
\label{sec:ode-resnet}

In this section, we briefly describe recent mathematical representation of deep Residual Neural Network (ResNet); for a comprehensive introduction see, e.g., \cite{chang2017multi,haber2017stable,lu2017beyond}. We outline the most important part of deep ResNet which is the forward propagation. For notation convenience, we stack the training features and target row-wise into matrices $\bX_0=[G^1, G^2,..., G^s]^T\in \mathcal{R}^{s\times n}$ and $\bA=[A^1, A^2,...,A^s]\in \mathcal{R}^{s\in N}$.
We consider a simplified version of ResNet model that has been successful in classifying images. The input values of forward propagation in the ResNet is given by,
\begin{eqnarray}
\label{eqn:forward}
\bX_{t+1} = \bX_t + \sigma(\bX_t\bW_t + b_t) \,\,\, t=0,....,N-1,
\end{eqnarray}
where $N$ is the number of layers in the network architectures, $X_0\in \mathcal{R}^{s\times n}$ is the initial input value. This propagation is parametrized by the nonlinear activation function $\sigma: \mathcal{R}^{s\times n}\rightarrow \mathcal{R}^{s\times n}$ and affine transformations represented by their weights, $\bW_0,..., \bW_{N-1}\in \mathcal{R}^{n\times n}$, and bias $b_0,...,b_{N-1}\in R^{1\times n}$. The values $\bX_t$ are called hidden layers and $\bX_N$ is the final output layer. The activation function is applied element-wise and is typically smooth and non-decreasing. Two commonly used examples are hyperbolic tangent (tanh) and the Rectified Linear Unit (ReLU) activations.
For simplicity, we only consider the ReLU activation in our model,
\begin{equation}
\sigma_{ReLU} = max(0, \bX).
\end{equation}
The final output layer predict the values using the hypothesis function $h(\bX)$. For our problem, we assume the spectral function (output of the network) satisfy multinomial distributions so that we can use the softmax function in the output layer,
\begin{eqnarray}
h(\bX) = \frac{exp(\bX)}{exp(\bX)\be_m},
\end{eqnarray}
where $\be_m\in \mathcal{R}^{m}$ denotes the $m-$dimensional vector of all ones. 

The learning problem is to estimate the parameters of the forward propagation so that the deep ResNet can accurately approximates the training data set. This learning process can be solved by the following optimization problem
\begin{eqnarray}
\min L(\tilde{\bA}, \bA) + \lambda R(\bW,b),
\end{eqnarray}
where the loss function $L(\tilde{\bA}, \bA)=1/2\|\tilde{\bA} - \bA\|^2_F$ is the sum of squared difference and the convex regularizer $R$ penalizes undesirable parameters and can prevents overfitting.

\subsection{Adams-Bashforth Scheme}
\label{sec:ab}
Much recent work has motivated us to view the ResNet as a dynamic system \cite{lu2017beyond,haber2017stable}. There is significant work on connecting numerical ODEs and deep neural networks. The work \cite{chang2017multi} adopt the dynamical systems point-of-view and analyze the lesioning properties of ResNet both theoretically and experimentally. The effort \cite{chen2018neural} introduced the ODE-net that can interpret and solve the ResNet using ODE solver, which provides memory efficiency for deep ResNet. In this article, motivated by the previous work, we propose a new Adams-Bashforth ResNet architecture for the analytic continuation problem. 

\begin{figure}[h!]
\includegraphics[scale=0.5]{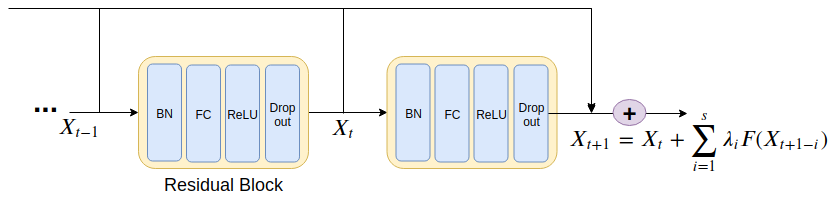}
\caption{Two step neural network}
\label{fig:2stepresnet}
\end{figure}

The forward propagation of~\eqref{eqn:forward} can be considered as the forward Euler discretization of the initial value ODE given by
\begin{eqnarray}
\label{eqn:ode-net}
\dot{\bX}(t) = \bFF(\bX(t),\bW(t),b(t)),\,\, \bX(0) = \bX_0,\,\, 0\leq t\leq T,
\end{eqnarray}
where time $t$ corresponds to the direction from input to output, $\bX(0)$ is the initial input feature, and $\bX(T)$ is the output of the network. Thus, the problem of learning the network parameters, $\bW$ and $b$, is equivalent to solving a parameter estimating problem or optimal control problem involving the ODE in~\eqref{eqn:ode-net}. Note that the time step size $\Delta t$ in the fully discretized ODE $\frac{\bX_{t+1} - 
\bX_t}{\Delta t} = \bFF(\bX_{t})$, is implicitly absorbed by the residual module in the original formulation of ResNet~\eqref{eqn:forward}. Instead, we intend to use a multistep Adams-Bashforth (AB) method to discretize~\eqref{eqn:ode-net}. As mentioned before, the standard ResNet can be considered as the forward Euler discretization, whereas multistep AB method has higher accuracy in numerical methods of ODE \cite{ascher1998computer}. The fully discretized schemd is shown in Fig.~\ref{fig:2stepresnet} and
\begin{eqnarray}
\bX_{t+s} = \bX_{t+s-1} + \Delta t\sum_{i=1}^s\lambda_i\bFF(\bX_{t+s-i}),
\end{eqnarray}
where $\sum_{i=1}^s\lambda_i = 1$. The formula can be derived from Taylor's theorem. As an example, we use two step method (AB2) to illustrate, i.e.,
\begin{eqnarray}
\bX_{t+1} &=& \bX(t)+\Delta t((1-\lambda)\dot{\bX}(t)+\lambda(\dot{\bX}(t)-\Delta t\ddot{\bX}(t) + \mathcal{O}(\Delta t^2))) \nonumber\\
&=& \bX(t)+\Delta t \dot{\bX}(t)-\lambda\Delta t^2\ddot{\bX}(t) + \mathcal{O}(\Delta t^3).
\end{eqnarray}
Then applying Taylor expansion on the true solution, i.e.,
\begin{eqnarray}
\bX(t+1) = \bX(t)+\Delta t\dot{\bX}(t)+\frac{1}{2}\Delta t^2\ddot{\bX}(t) + \mathcal{O}(\Delta t^3),
\end{eqnarray}
The numerical scheme associated to the AB2 and AB3 is the following
\begin{eqnarray}
\bX_{t+1} = \bX_{t} + \frac{3}{2}\bFF(\bX_t,\bW_t,b_t) - \frac{1}{2}\bFF(\bX_{t-1},\bW_{t-1}, b_{t-1}),
\end{eqnarray}
\begin{eqnarray}
\bX_{t+1} = \bX_{t} + \frac{23}{12}\bFF(\bX_t,\bW_t,b_t) - \frac{4}{3}\bFF(\bX_{t-1},\bW_{t-1}, b_{t-1})+\frac{5}{12}\bFF(\bX_{t-2},\bW_{t-2}, b_{t-2}).
\end{eqnarray}
The AB2 method has second order $\mathcal{O}(\Delta t^2)$ accuracy. Standard ResNet is considered a AB1 method which has first order $\mathcal{O}(\Delta t)$ accuracy. According to the stability analysis of linear multistep explicit methods, the AB3 method is strongly stable while AB2 and AB1 is conditional stable. This stability property drives us to apply the AB method to obtain a more robust deep network architectures that can provide a model with better performance for noisy data. The family of linear multistep method is large. To shorten the discussion in this work, we focus on the AB2 and AB3 method in our numerical tests.

\section{Numerical Experiment}
\label{sec:numerical}

\subsection{Dataset}
\label{sec:dataset}
In this section, we present the numerical results from our new model. The training data can be collected from experimental measurements or simulated according to a theoretical model. In this work, we choose to simulate spectral density functions that always have a quasiparticle peak close to $\omega=0$, as often encountered when considering correlated metals. In the data generation, the spectral densities $A(\omega)$ are defined as a sum of uncorrelated Gaussian distributions:
\begin{eqnarray}
A^i(\omega) = \frac{1}{R_i}\sum_{k=0}^{R_i} exp\biggl(-\frac{(\omega-\mu_k)^2}{2\sigma^2_k}\biggr),
\end{eqnarray}
where the frequencies $\omega\in [-10, 10]$, the centers of the peaks $\mu_k\in [-5, 5]$, the number of Gaussian distributions $R_i\in [1,...,21]$, and $\sigma_k\in [0.1, 1]$. Parameters $R_i, \mu_k, \sigma_k$ are uniformly sampled over the above-mentioned ranges. The Green's functions are then computed by Eq.~\eqref{eq:AC}. The discretization of the Green's function is generally over $\mathcal{O}(10^3)$. The amount of data necessary to approximate a function grows exponential with the number of dimensions. To reduce the effect of the curse of dimensionality, we use the orthogonal Legendre polynomials to represent the Green's function data which can facilitates the learning process of the model. The compact representation is given by $G(\tau) = \sum (2l+1)G_lP_l(2\frac{\tau}{\beta}-1)$, where $P_l$ are the legendre polynomials. In the experiments, 64 basis are used to ensure the accurate approximation of the data, with similar strategies found in \cite{fournier2018artificial}. Three noise levels $\epsilon=10^{-5}, 10^{-3}, 10^{-2}$ are added to the dataset, such that, $G^{train} = G+\epsilon$. 
\begin{figure}[h!]
    \includegraphics[scale=0.3]{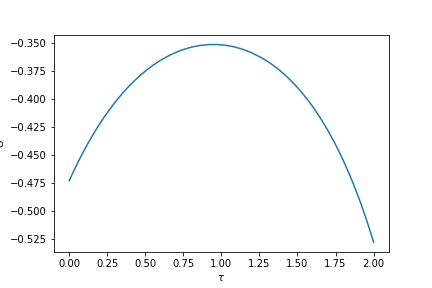}
    \includegraphics[scale=0.3]{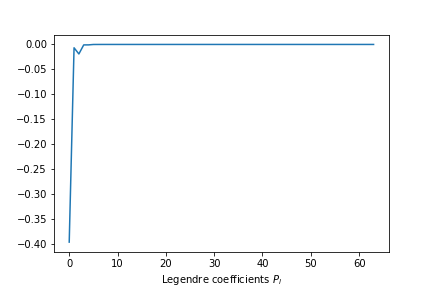}
    \includegraphics[scale=0.3]{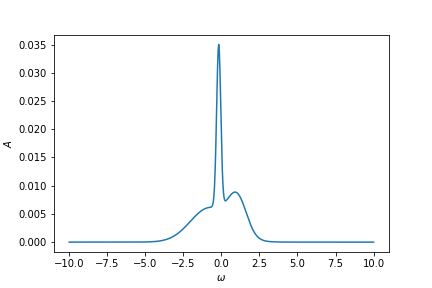}
    \caption{One data sample from the training set $G(\tau)$ (left), Legendre representation $G_l$ (mid), and target spectral density $A(\omega)$ (right)}
\label{fig:sample}
\end{figure}

\subsection{Numerical Results}
\label{sec:num-res}
The network architecture of our model, shown in Fig.~\ref{fig:2stepresnet}, consists of an input layer connected to a residual block, followed by eight repetitions of the residual block. For each residual block, the first layer is a batch normalization followed by a fully collected dense layer with ReLU activation. Then follows a dropout layer that helps to avoid overfitting issueds by randomly dropping units. The output is computed using a softmax layer, which ensure the similarity to a probability density function. The training is performed
on a dataset of size 100,000 with validation and test sets, both of size 1000, used in our numerical experiment. The code implementation is based on PyTorch where the Adams optimizer and the KLD loss function were used. 

\begin{figure}[h!]
\centering
\includegraphics[scale=0.5]{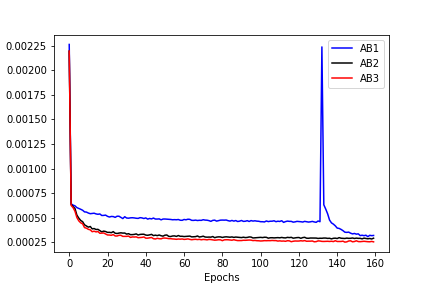}
\caption{The training performance from AB1-ResNet, AB2-ResNet, and AB3-ResNet structure with data noise $10^{-2}$}
\label{fig:training}
\end{figure}

We have investigated AB-ResNet approaches to improving the robustness of our model against noisy data. As mentioned before, the AB1 (ResNet) and AB2 method is conditional stable whereas AB3 is strongly stable. In order to study the stability of the network achitecture nuemrically, we trained each model on the dataset with different magnitude of noisy, i.e., $10^{-5}$, $10^{-3}$, and $10^{-2}$. Fig.~\ref{fig:training} shows the training performance from each network and, as expected, the AB3 network has a better learning behavior than AB1 and AB2. 

\begin{figure}[h!]
\includegraphics[scale=0.3]{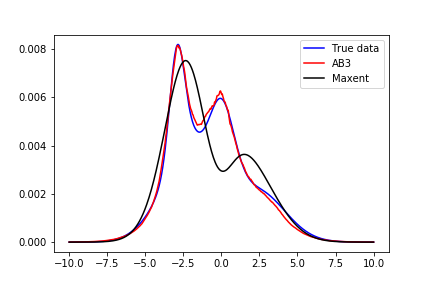}
\includegraphics[scale=0.3]{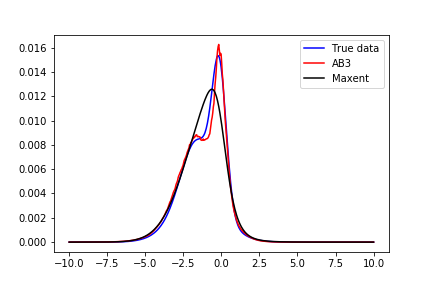}
\includegraphics[scale=0.3]{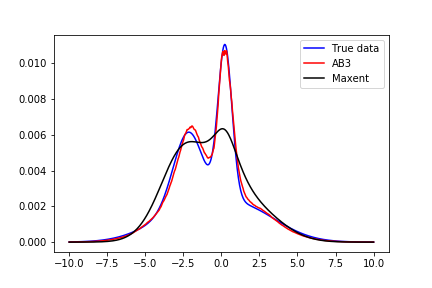}\\
\includegraphics[scale=0.3]{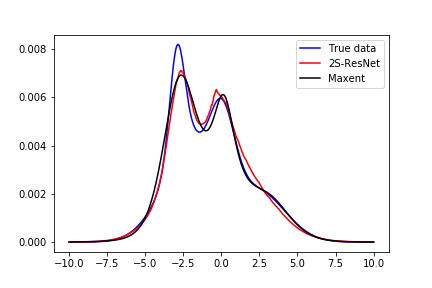}
\includegraphics[scale=0.3]{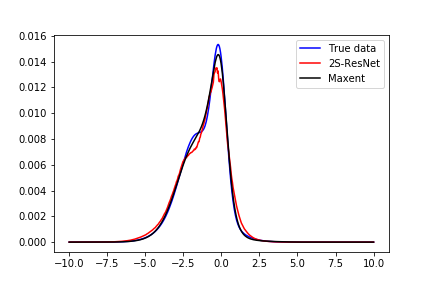}
\includegraphics[scale=0.3]{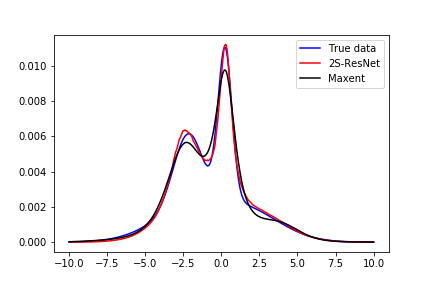}
\caption{The predicted spectral density function $A(\omega)$ from AB3-ResNet and Maxent (dark line). The top row are the results from dataset with noise level $10^{-2}$, the bottom row results obtained from the dataset under noise level $10^{-3}$}
\label{fig:comparemaxent}
\end{figure}

\begin{figure}[h!]
\includegraphics[scale=0.3]{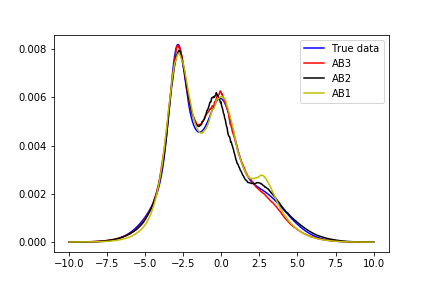}
\includegraphics[scale=0.3]{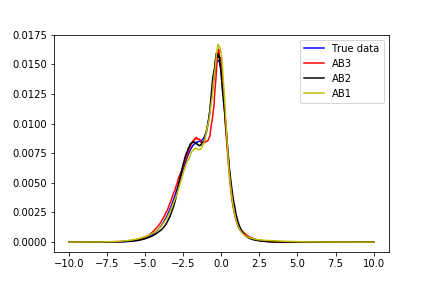}
\includegraphics[scale=0.3]{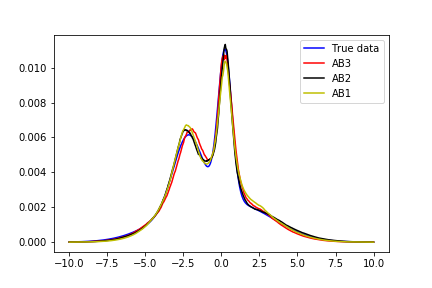}
\caption{The comparison of predicted spectral function between 2S-ResNet (red) and standard Neural Net (dark)}
\label{fig:compareNN}
\end{figure}

Fig.~\ref{fig:comparemaxent} provides a qualitative comparison of the results of our AB-ResNet method and the MaxEnt method where we plot three samples from test set for illustration purposes. In these examples, both methods predict $A(\omega)$ accurately for the lowest level of noise. However, at noise $\epsilon=10^{-2}$, MaxEnt is not able to recover the peaks in the predicted spectral function. While in the case of AB-ResNet, our model is able to correctly identify most peaks. Hence, it clearly shows that our AB-ResNet model generates better results compared to the classical MaxEnt.  Fig.~\ref{fig:compareNN} shows the comparison of the prediction between each AB network model from three different samples. The average mean absolute error on the test dataset are $6.8e-4, 3.8e-4, 2.6e-4$ for AB1, AB2 and AB3, respectively. This is consistent with the numerical ODE. That is, higher step method provides higher accuracy results. Then, we studied the computational efficiency of our model compared to MaxEnt. AB-ResNet model allows a direct mapping between Green's function and the spectral densities. In contrast, the MaxEnt method is an iterative method which requires generating trail functions until convergence is reached. For the computation cost, the CPU time for AB-ResNet model is $~\mathcal{O}(10)$ second while for MaxEnt is $~\mathcal{O}(10^3)$ second. So, the new model is more computationally efficient than compared to MaxEnt method. 

\section{Conclusions}
\label{sec:conclu}
In summary, we have developed the Adams-Bashforth ResNet that solves the kernel inversion with noisy data for the analytic continuation problem. The numerical experiments show that our AB-ResNet model can recover the spectral function with an accuracy similar to that of the commonly used maximum entropy approach under low levels of noise. The new model gives much better results than MaxEnt under high levels of noise at a fraction of its computational cost. Adding more training data and using larger step network architecture could further improve the model performance. Other inverse problem can apply our model the same way given the great representative capacity of deep AB-ResNet.

Some future work should consider the limitations of the proposed model. One main drawback of the method is that the model is learned for a particular inverse temperature, i.e., $\beta=2$, whereas the MaxEnt method can provide it as a parameter. So, the MaxEnt method has the generality with respect to different $\beta$. To extend our model to arbitrary values of $\beta$, we can train a separate network for each parameter. Another approach for this issue would be to add $\beta$ as an input parameter to the model and train it on a large collection of dataset. These approaches can improve the robustness of our model with respect to the inverse temperature. Another direction of improving the model would be using implicit scheme in the neural network architecture, since implicit method is more stable than explicit method. It might have a better convergence for the learning process, which needs to be future investigated.


\subsubsection*{Acknowledgments}
This material is based upon work supported in part by: the Scientific Discovery through Advanced Computing (SciDAC) program, U.S. Department of Energy, Basic Energy Sciences, Division of Materials Sciences and Engineering; the U.S. Department of Energy, Office of Science, Early Career Research Program under 
award number ERKJ314; U.S. Department of Energy, Office of Advanced Scientific Computing Research under award numbers ERKJ331 and ERKJ345; 
the National Science Foundation, Division of Mathematical Sciences, Computational Mathematics program under contract number DMS1620280;
 and by the Laboratory Directed Research and Development program at the Oak Ridge National Laboratory, which is operated by UT-Battelle, LLC., for the U.S. Department of Energy under contract DE-AC05-00OR22725.

\begin{thebibliography}{10}

\bibitem{arsenault2016projected}
Louis-Francois Arsenault, Richard Neuberg, Lauren~A Hannah, and Andrew~J
  Millis.
\newblock Projected regression methods for inverting fredholm integrals:
  Formalism and application to analytical continuation.
\newblock {\em arXiv preprint arXiv:1612.04895}, 2016.

\bibitem{ascher1998computer}
Uri~M Ascher and Linda~R Petzold.
\newblock {\em Computer methods for ordinary differential equations and
  differential-algebraic equations}, volume~61.
\newblock Siam, 1998.

\bibitem{Bao:2016ca}
F~Bao, Y~Tang, M.~Summers, G~Zhang, C~Webster, V~Scarola, and T.~A. Maier.
\newblock {Fast and efficient stochastic optimization for analytic
  continuation}.
\newblock {\em Physical Review B}, 94(12):125149, September 2016.

\bibitem{Beach:04}
K.~S.~D. Beach.
\newblock Identifying the maximum entropy method as a special limit of
  stochastic analytic continuation.
\newblock {\em preprint arXiv:0403055 year = {2004}}.

\bibitem{beck2017machine}
Christian Beck, E~Weinan, and Arnulf Jentzen.
\newblock Machine learning approximation algorithms for high-dimensional fully
  nonlinear partial differential equations and second-order backward stochastic
  differential equations.
\newblock {\em Journal of Nonlinear Science}, pages 1--57, 2017.

\bibitem{chang2018reversible}
Bo~Chang, Lili Meng, Eldad Haber, Lars Ruthotto, David Begert, and Elliot
  Holtham.
\newblock Reversible architectures for arbitrarily deep residual neural
  networks.
\newblock In {\em Thirty-Second AAAI Conference on Artificial Intelligence},
  2018.

\bibitem{chang2017multi}
Bo~Chang, Lili Meng, Eldad Haber, Frederick Tung, and David Begert.
\newblock Multi-level residual networks from dynamical systems view.
\newblock {\em arXiv preprint arXiv:1710.10348}, 2017.

\bibitem{chen2018neural}
Tian~Qi Chen, Yulia Rubanova, Jesse Bettencourt, and David~K Duvenaud.
\newblock Neural ordinary differential equations.
\newblock In {\em Advances in Neural Information Processing Systems}, pages
  6571--6583, 2018.

\bibitem{fournier2018artificial}
Romain Fournier, Lei Wang, Oleg~V Yazyev, and QuanSheng Wu.
\newblock An artificial neural network approach to the analytic continuation
  problem.
\newblock {\em arXiv preprint arXiv:1810.00913}, 2018.

\bibitem{Fuchs:2010it}
Sebastian Fuchs, Thomas Pruschke, and Mark Jarrell.
\newblock {Analytic continuation of quantum Monte Carlo data by stochastic
  analytical inference}.
\newblock {\em Physical Review E}, 81(5):056701, May 2010.

\bibitem{fuchs2010analytic}
Sebastian Fuchs, Thomas Pruschke, and Mark Jarrell.
\newblock Analytic continuation of quantum monte carlo data by stochastic
  analytical inference.
\newblock {\em Physical Review E}, 81(5):056701, 2010.

\bibitem{Gull:1984jk}
S~F Gull and J~Skilling.
\newblock {Maximum entropy method in image processing}.
\newblock {\em IEE Proceedings F (Communications, Radar and Signal
  Processing)}, 131(6):646--659, October 1984.

\bibitem{haber2017stable}
Eldad Haber and Lars Ruthotto.
\newblock Stable architectures for deep neural networks.
\newblock {\em Inverse Problems}, 34(1):014004, 2017.

\bibitem{he2016deep}
Kaiming He, Xiangyu Zhang, Shaoqing Ren, and Jian Sun.
\newblock Deep residual learning for image recognition.
\newblock In {\em Proceedings of the IEEE conference on computer vision and
  pattern recognition}, pages 770--778, 2016.

\bibitem{Jarrell:1996uo}
M.~Jarrell and J.~Gubernatis.
\newblock {Bayesian Inference and the Analytic Continuation of Imaginary-Time
  Quantum Monte Carlo Data}.
\newblock {\em Physics Reports}, 269:133--195, 1996.

\bibitem{jarrell1996bayesian}
Mark Jarrell and James~E Gubernatis.
\newblock Bayesian inference and the analytic continuation of imaginary-time
  quantum monte carlo data.
\newblock {\em Physics Reports}, 269(3):133--195, 1996.

\bibitem{levy2017implementation}
Ryan Levy, JPF LeBlanc, and Emanuel Gull.
\newblock Implementation of the maximum entropy method for analytic
  continuation.
\newblock {\em Computer Physics Communications}, 215:149--155, 2017.

\bibitem{lite19jmlr}
Qianxiao Li, Cheng Tai, and Weinan E.
\newblock Stochastic modified equations and dynamics of stochastic gradient
  algorithms {I:} mathematical foundations.
\newblock {\em Journal of Machine Learning Research}, 20:40:1--40:47, 2019.

\bibitem{lin2018resnet}
Hongzhou Lin and Stefanie Jegelka.
\newblock Resnet with one-neuron hidden layers is a universal approximator.
\newblock In {\em Advances in Neural Information Processing Systems}, pages
  6169--6178, 2018.

\bibitem{long18a}
Zichao Long, Yiping Lu, Xianzhong Ma, and Bin Dong.
\newblock {PDE}-net: Learning {PDE}s from data.
\newblock In {\em Proceedings of the 35th International Conference on Machine
  Learning}, pages 3208--3216, 2018.

\bibitem{lu2017beyond}
Yiping Lu, Aoxiao Zhong, Quanzheng Li, and Bin Dong.
\newblock Beyond finite layer neural networks: Bridging deep architectures and
  numerical differential equations.
\newblock In {\em Proceedings of the 35th International Conference on Machine
  Learning}, pages 3282--3291, 2018.

\bibitem{ma2018model}
Chao Ma, Jianchun Wang, et~al.
\newblock Model reduction with memory and the machine learning of dynamical
  systems.
\newblock {\em arXiv preprint arXiv:1808.04258}, 2018.

\bibitem{Mishchenko:2000co}
A~S Mishchenko, N~V Prokof{\textquoteright}ev, A~Sakamoto, and B~V Svistunov.
\newblock {Diagrammatic quantum Monte Carlo study of the Fr{\"o}hlich polaron}.
\newblock {\em Physical Review B}, 62(10):6317--6336, September 2000.

\bibitem{Prokofev:2013hi}
N~V Prokof{\textquoteright}ev and B~V Svistunov.
\newblock {Spectral analysis by the method of consistent constraints}.
\newblock {\em Jetp Lett.}, 97(11):649--653, August 2013.

\bibitem{Sandvik:1998jr}
Anders Sandvik.
\newblock {Stochastic method for analytic continuation of quantum Monte Carlo
  data}.
\newblock {\em Physical Review B}, 57(17):10287--10290, May 1998.

\bibitem{Silver:1990eb}
R~N Silver, J~E Gubernatis, D~S Sivia, and M.~Jarrell.
\newblock {Spectral densities of the symmetric Anderson model}.
\newblock {\em Physical Review Letters}, 65(4):496--499, July 1990.

\bibitem{wang2018deep}
Bao Wang, Xiyang Luo, Zhen Li, Wei Zhu, Zuoqiang Shi, and Stanley Osher.
\newblock Deep neural nets with interpolating function as output activation.
\newblock In {\em Advances in Neural Information Processing Systems}, pages
  743--753, 2018.

\bibitem{wang2018exponential}
Qingcan Wang et~al.
\newblock Exponential convergence of the deep neural network approximation for
  analytic functions.
\newblock {\em arXiv preprint arXiv:1807.00297}, 2018.

\bibitem{wume18nips}
Lei Wu, Chao Ma, and Weinan E.
\newblock How {SGD} selects the global minima in over-parameterized learning:
  {A} dynamical stability perspective.
\newblock In {\em Advances in Neural Information Processing Systems 31: Annual
  Conference on Neural Information Processing Systems 2018, NeurIPS 2018, 3-8
  December 2018, Montr{\'{e}}al, Canada.}, pages 8289--8298, 2018.

\bibitem{yoon2018analytic}
Hongkee Yoon, Jae-Hoon Sim, and Myung~Joon Han.
\newblock Analytic continuation via domain knowledge free machine learning.
\newblock {\em Physical Review B}, 98(24):245101, 2018.

\end{thebibliography}


%
%
%
%
%

\end{document}